\documentclass[pra,twocolumn,twoside]{revtex4}

\usepackage{bm} % bold math
\usepackage{graphics}
\usepackage{amssymb}
\usepackage{hyperref}

\begin{document}

\title{Instantaneous coherent destruction of tunneling and fast quantum state preparation for strongly pulsed spin qubits in diamond}
\date{\today}

\author{Martijn Wubs}
\email{mwubs@fotonik.dtu.dk}
\homepage{http://www.mortensen-lab.org/wubs}
\address{DTU Fotonik, Technical University of Denmark, DK-2800 Kgs. Lyngby, Denmark}

\begin{abstract}
Qubits driven by resonant strong pulses are studied and a parameter regime is explored in which the dynamics can be solved in closed form. Instantaneous coherent destruction of tunneling can be seen for longer pulses, whereas shorter pulses allow a fast preparation of the qubit state. Results are compared with recent experiments of pulsed nitrogen-vacancy center spin qubits in  diamond.
\end{abstract}


\maketitle

\section{Introduction}
\label{intro}

Quantum systems with well-isolated two-level subsystems that can be addressed independently, with states that can be initialized and manipulated, enjoy much interest nowadays because of possible use as quantum bits in quantum information processing. Quantum-state manipulation is usually described by time-dependent Hamiltonians, and the time-dependent (driving) part of the Hamiltonian can either be weak or strong. Weak driving is the default situation, but here the focus will be on strongly driven quantum systems.

In the weak-driving regime, the effect of faster and larger-amplitude driving may be the same as of slow small-amplitude driving, as only the area under the pulse is important. This is well known for example for $\pi$-pulses, which are used to invert the population of a qubit. Since decoherence sets the time scale within which quantum information processing should occur, it is then a good idea to speed up quantum state manipulation by making pulses shorter, with an amplitude of the pulses that grows concomitantly to keep the same pulse area. This strategy works until the peak amplitude grows so large that the driving can no longer be considered weak, and the dynamics becomes more complex.

Nevertheless, it may be a good idea to go to the strong-driving regime to seek new ways to manipulate qubits fast compared to the decoherence time~\cite{Fuchs2009a}. The strong-driving limit has not been studied extensively in a quantum information perspective. Important was the  discovery of dynamical decoupling of open quantum systems from their environments~\cite{Viola1999a,FonsecaRomero2005a}. The recent advent of novel types of qubits with strong couplings to cavities and external fields naturally drives the interest in strongly driven quantum systems, for example in the context of amplitude spectroscopy of superconducting qubits~\cite{Oliver2005a,Sillanpaa2006a,Oliver2009a,Shevchenko2009a} and optomechanical systems~\cite{Heinrich2010a}.

Actually, the interest in strongly driven quantum systems dates back before quantum information theory. The strong-coupling regime was studied for example in chemical physics, in order to control a molecular reaction by intense laser pulses~\cite{Holthaus1992a,Breuer1993a}. Many strongly driven quantum systems exhibit common phenomena for which there is no analogy in the weak-driving limit. One such universal phenomenon is {\em coherent destruction of tunneling} (CDT), which was discovered theoretically and explained in 1991 by  Grossmann, Dittrich, Jung, and H{\"a}nggi~\cite{Grossmann1991a,Grossmann1992a}. For a qubit, it is the phenomenon  that a tunneling between the two quantum states is brought to a standstill by strong driving, but only for specific values of the quotient of driving amplitude and frequency (details given below). For other parameter values, the tunneling amplitude is  renormalized, though not `destructed', due to the driving.  An explanation of CDT  in terms of destructive interference of multiple Landau-Zener-St{\" u}ckelberg transitions was given by Kayanuma soon after~\cite{Kayanuma1994a}, see also~\cite{Saito2008a,Asshab2007a}.

CDT is currently actively studied in several subfields of physics. Here I only mention a few recent developments.
The recent single-particle tunneling experiment on strongly driven metastable argon in optical double-well potentials~\cite{Kierig2008a} is a very direct observation of CDT, and close in spirit to the original theoretical papers~\cite{Grossmann1991a,Grossmann1992a}.
Very interesting are recent many-body generalizations of CDT, both theoretical predictions~\cite{Eckardt2005a,Eckardt2007a,Luo2007a,Gong2009a,Zueco2009a} and experimental demonstrations~\cite{Lignier2007a,Sias2008a}.  For example, for a driven Bose-Einstein condensate in a double well, many-body CDT phenomena are predicted that sensitively depend on the number of particles that have already tunneled~\cite{Gong2009a}. Furthermore, for  Bose-Einstein condensates in shaken optical lattices, it was predicted~\cite{Eckardt2005a} and shown experimentally~\cite{Lignier2007a} that  the shaking-induced renormalization of the tunneling parameter enables the switching between the superfluid and the insulator state. Interesting is also the recent proposal for routing of quantum information by employing coherent destruction of tunneling in chains of qubits~\cite{Zueco2009a}. Moreover, coherent destruction of tunneling is also seen in optics, in particular in coupled-waveguide structures~\cite{Vorobeichik2003a,Luo2007a,DellaValle2007a,Szameit2009a}, for which the coupled spatial wave equations have the same mathematical form as the temporal equations of motion for quantum systems exhibiting CDT~\cite{Longhi2005a}. Further work on CDT can be found in the reviews~\cite{Grifoni1998a,Platero2004a,Kohler2005a}.

The standard system for observing CDT is a harmonically driven two-level system. With fast manipulation of qubits in mind, here we consider driving by  pulses instead, allowing for several types of pulse envelopes, and discuss the possibility of observing and exploiting CDT for quantum information processing in this more complex case. Related work on pulsed systems can be found in~\cite{Holthaus1992a,Breuer1993a,Oliver2005a,Kleinekathoefer2006a}.

We then apply our analysis to  nitrogen-vacancy (NV) center spin qubits in diamond strongly driven by microwave pulses. This is a very promising type of solid-state qubit for several reasons~\cite{Wrachtrup2006a}. For example, its coherence time is long even at room temperature, and its state can be initialized by optical pumping. And one can transfer its state to even longer-lived nuclear spins~\cite{Dutt2007a,Smeltzer2009a}.  NV center spin qubits may become building blocks of a quantum repeater~\cite{Childress2006a}. Achieving coherent coupling between NV centers and superconducting flux qubits is another exciting possibility~\cite{Marcos2010a}.

Although our results are general, the very recent measurements by Fuchs {\em et al.}~\cite{Fuchs2009a} have shown a great level of control of the fast and strong microwave pulses by which NV center spin qubits can be prepared in desired quantum states, and were the main motivation for carrying out the research presented here.
We discuss under what circumstances CDT in NV center spin qubits could be observed, and present analytical results that can be of help to prepare final quantum states using short resonant pulses.

The structure of this article is as follows.  We introduce our model in Sec.~\ref{Sec:model} with dynamics and some approximations. This is followed by the exploration of CDT phenomena in strongly pulsed qubits in Sec.~\ref{Sec:ICDT}. The approximate treatments and numerically exact results are compared in Sec.~\ref{Sec:numcomp}. The choice of optimized pulses is briefly discussed in Sec. \ref{pulseshaping}. The application of the analysis to nitrogen-vacancy  centers in diamond can be found in Sec.~\ref{Sec:diamond}, before the conclusions in Sec.~\ref{Sec:conc}.

\section{Qubit driven by a strong pulse: model, dynamics, and approximations}
\label{Sec:model}

\subsection{Model and equations of motion}
Consider a  two-level system with states $\{ |0\rangle,|1\rangle\}$ driven by a pulse, as described by the Hamiltonian
\begin{eqnarray}\label{CDTHamiltonianstandard}
H(t) & = & h_{z}(t)\sigma_{z} + h_{x}\sigma_{x} = \frac{\hbar}{2}[ \omega_{0} + V(t)]{\bm \sigma}_{z} + \hbar\Delta {\bm \sigma}_{x} \nonumber \\
& = & \hbar\left( \begin{array}{cc} \omega_{0}/2 + V(t)/2 & \Delta \\
\Delta &  -\omega_{0}/2 - V(t)/2  \end{array} \right).
\end{eqnarray}
where   $\omega_{0}$ is the energy difference  and $\Delta$  the interaction strength between the two states of the undriven qubit, and $V(t) = A(t) \cos (\omega t)$ is the driving field with frequency $\omega$ and pulse envelope $A(t)$. The $\sigma_{x,z}$ are Pauli operators, with $\sigma_{x}$ defined as $|{0}\rangle\langle{1}| + |{1}\rangle\langle{0}|$ and $\sigma_{z}=|{1}\rangle\langle{1}| - |{0}\rangle\langle{0}|$.
For vanishing driving amplitude or as long as $A(t)$ is still identically zero after the initial time $t_{0}$, the  Hamiltonian~(\ref{CDTHamiltonianstandard}) is constant and describes simple tunneling dynamics with sinusoidal population exchange between the two levels at a frequency $\sqrt{\omega_{0}^{2}+4\Delta^2}$. For $\omega_{0}=0$, the excited-state population $P_{1}(t)$ varies between 0 and 1, but if $\Delta/\omega_{0} \ll 1$, then the excited-state population stays negligible before the driving is switched on.
For $A(t)=A$ constant, the driving is harmonic, and
Eq.~(\ref{CDTHamiltonianstandard}) in the particular case of constant-amplitude driving will be referred to as the CDT Hamiltonian in standard form, or standard CDT Hamiltonian. With the goal in mind of fast interactions with qubits, in the following we consider instead the qubit dynamics under strong driving with short pulses.

With the help of the time evolution operator $U_{0}(t) = \exp\left[ - (i/\hbar)\sigma_{z}\int^{t}\mbox{d}\tau h_{z}(\tau)   \right]$, an interaction picture can be defined in which the Hamiltonian $~\tilde H(t)$ is given by $ U_{0}^{\dag}(t) \Delta \sigma_{x}U_{0}(t) $. In matrix representation for the state $|\tilde\psi(t)\rangle = c_{1}(t)|1\rangle+ c_{0}(t)|0\rangle$, the coupled equations of motion for the two coefficients become
\begin{equation}\label{H2standardmatrix}
i \hbar \left( \begin{array}{c} \dot c_{1} \\ \dot c_{0} \end{array}\right) =
\left( \begin{array}{cc} 0  & \Delta(t) \\
\Delta^{*}(t)  & 0 \end{array}\right) \left( \begin{array}{c}  c_{1} \\  c_{0} \end{array}\right),
\end{equation}
where the dots denote time derivatives and the time-dependent interaction is defined as $\Delta(t)\equiv \Delta  \exp\left[  2i/\hbar\int^{t}\mbox{d}\tau h_{z}(\tau)   \right]$. These coupled equations of motion can be solved numerically without further ado, but some striking features of the dynamics can be understood after only a little more analysis.

\subsection{Rotating-wave approximation}\label{Sec:RWA}
Driving with a pulse adds some complexity to the dynamics as compared to harmonic driving. To keep the presentation as simple as possible, let us first choose a very convenient pulse shape, obtain some analytical results, and then argue that the results can be generalized to other pulses. We choose the simple envelope function $A(t)= A p(t)$,  with maximum amplitude $A$ and dimensionless pulse shape function $p(t) = \exp(- |t|/\tau_{p})$, a two-sided  exponential with decay rate $\tau_{p}^{-1}$. In this particular case, the integral in the exponents of Eq.~(\ref{H2standardmatrix}) exactly becomes
\begin{equation}\label{hzexact}
  2i/\hbar\int^{t}\mbox{d}\tau h_{z}(\tau)   = i\omega_{0}t +i \cos(\chi) A(t)\sin(\omega t \mp \chi)/\omega,
\end{equation}
with $\chi \equiv \arctan[1/(\omega\tau_{p})]$ and $\mp = -t/|t|$.
The amplitude  $A$ is reduced by a factor $\cos\chi$, and $\chi$ also shows up as a phase factor. Our convenient pulse shape enables us to proceed as in the analysis of CDT for harmonic driving by using the mathematical identity known as the Jacobi-Anger expansion,
\begin{equation}
e^{i x \sin \alpha} = \sum_{n = -\infty}^{\infty}J_{n}(x) \,e^{i n \alpha},
\end{equation}
where $J_{n}$ denotes as usual a Bessel function of the first kind. The upper-right matrix element $\tilde H_{10}$ in Eq.~(\ref{H2standardmatrix}) then becomes
\begin{equation}\label{H10}
\Delta\sum_{n=-\infty}^{\infty} J_{n}\left[\cos(\chi) A(t)/\omega\right]e^{\mp i n \chi } \,e^{i (\omega_{0} + n \omega) t},
\end{equation}
where the $\mp$ signifies that the phases in $\exp(\mp i n \chi)$ flip sign at $t=0$. We have not made approximations yet. We will make one now, based on  the widely different time dependencies of the terms in the expansion~(\ref{H10}).    One can choose the driving frequency $\omega$ such that there is an $n$-photon resonance, {\em i.e.}  $\omega_{0}+n_{\rm res} \omega=0$ for some integer $n_{\rm res}$, and zero- and one-photon resonances ($n_{\rm res}=0,-1$) are most common.  For harmonic driving this makes the $n_{\rm res}$-term stationary, while all others oscillate. All other terms can be neglected when even the slowest other terms,  the $(n_{\rm res}\pm 1)$-terms, average out on the interaction time scale $\Delta^{-1}$. This is a rotating-wave approximation (or RWA) and for harmonic driving it is valid if $ \omega \gg \Delta$.
Throughout this paper we will indeed make this assumption of high-frequency driving.

In our case of pulses, the situation is slightly more complex, since the Bessel-function coefficients have become time-dependent, including the resonant $n_{\rm res}$ term. For slowly-varying pulses, the almost-stationary term will vary much slower than all other terms, and it is intuitive that it is still a good approximation to keep only this term. For very short pulses containing only a few oscillations, the validity of the approximation of keeping only one time-dependent term is not so clear, and we will test it numerically below.
(A thorough discussion of the RWA of pulsed systems could be given using an adiabatic theorem for Floquet-Bloch states~\cite{Holthaus1992a,Kohler2005a}.)
The RWA dynamics is described by the equations of motion
\begin{equation}\label{doubleexRWA}
i  \left( \begin{array}{c} \dot c_{1} \\ \dot c_{0} \end{array}\right) \simeq
\Delta  J_{n_{\rm res}}\left[\cos(\chi) A(t)/\omega\right]\sigma_{x} \left( \begin{array}{c}  c_{1} \\  c_{0} \end{array}\right),
\end{equation}
where we got rid of phase factors by defining $\tilde c_{1} =  \exp[\pm i n_{\rm res} \chi/2] c_{1}$ and $\tilde c_{0} =  \exp[\mp i n_{\rm res} \chi/2] c_{0}$, and then leaving out the tildes~\cite{Wubs2005a}. As in the standard analysis of CDT, we see that due to the strong resonant driving, the interaction $\Delta$ is replaced by an effective interaction strength  $\Delta_{\rm eff} = \Delta  J_{n_{\rm res}}\left[\cos(\chi) A(t)/\omega\right]$. Since the Bessel functions are bounded by unity, the effective interaction is always smaller.

For harmonic driving, coherent destruction of tunneling is a  vanishing effective interaction, $\Delta_{\rm eff} = 0$. It occurs for those specific values of $A/\omega$ for which $J_{n_{\rm res}}(A/\omega)$  vanishes, and recall that $n_{\rm res}$ was fixed by our choice of resonant driving frequency $\omega$. Actually the name coherent destruction of tunneling is usually reserved for the case $n_{\rm res}=0$ and $\omega_{0}=0$, where it is the driving $\omega \gg \Delta$ that stops the tunneling that otherwise exists between the two undriven degenerate energy levels.
Here the name CDT will be more generally used for situations where $\Delta_{\rm eff} = 0$, even if for resonant driving with $n_{\rm res}\ne 0$, tunneling is also already suppressed without driving since $\omega_{0} \gg \Delta$.

The essential difference  for our pulsed driving is of course that the effective interaction $\Delta_{\rm eff}(t) = \Delta  J_{n_{\rm res}}\left[\cos(\chi) A(t)/\omega\right]$ has become a time-dependent quantity. This means that coherent destruction of tunneling will not occur at all times. Rather, as the pulse amplitude is varied from zero to a large maximal value $A \gg \omega$ and back, then $\Delta_{\rm eff}(t)$  vanishes at those instances in time at which the Bessel function vanishes. This phenomenon could be called instantaneous coherent destruction of tunneling, or ICDT, and will be studied in Sec.~\ref{Sec:ICDT}. Central result of this section is the equations of motion~(\ref{doubleexRWA}), obtained for the double-sided exponential pulse shape, with the RWA as the only approximation needed to get there.

\subsection{Slowly-varying envelope approximation}
Now let us see whether the results of Sec.~\ref{Sec:RWA}, which were obtained for a particular pulse shape, carry over to more general pulses. Recall that we did the integral $\int^{t}\mbox{d}\tau\,h_{z}(\tau)$ and the resulting Eq.~(\ref{hzexact})  that we obtained for the two-sided exponentially decaying pulse depended on a parameter $\chi=\arctan[1/(\omega \tau_{p})]$. Now if there are many oscillations per pulse, then $\omega \tau_{p}$ is much larger than unity and $\chi$ will almost vanish. But if we take the limit $\chi=0$ in the exact solution~(\ref{hzexact}), then we find, for $\omega \tau_{p} \rightarrow \infty$, that
\begin{equation}\label{hzexact}
h_{z}(t) = \frac{1}{2}\int^{t}\mbox{d}\tau\,\bigl\{ \omega_{0} + A(t)\cos(\omega t)\bigl\}\; \rightarrow \;\frac{\omega_{0}t}{2} + \frac{A(t)\sin(\omega t)}{2\omega}.
\end{equation}
Notice that there is a simpler way of obtaining the approximate result on  the right-hand side, namely by taking the amplitude $A(t)$ out of the time integral on the left-hand side as if it were a constant. This is a slowly-varying amplitude approximation (or SVAA). For the double-sided exponential pulse we will now derive the condition under which the RWA plus SVAA approximated dynamics is close to the exact dynamics, for a large but not infinitely large number of oscillations ($\omega\tau_{p} \gg 1$). We can make a  Taylor expansion to first order in $\chi$ of the left-hand side of Eq.~(\ref{hzexact}). This gives the approximate result on the right-hand side of Eq.~(\ref{hzexact}) plus a term $\pm A(t)\cos(\omega t)\chi/\omega$. We require the latter to be small at all times, which gives for the double-sided exponential pulses the following combined condition for the SVAA dynamics to be accurate:
\begin{equation}\label{SVAAcondition}
\omega \tau_{p}\gg 1 \qquad\mbox{and}\qquad A/\omega \ll \omega \tau_{p}.
\end{equation}
Clearly, the second condition on the scaled amplitude $A/\omega$ is only very weak once the first condition is satisfied.

We will now generalize our results to other pulse shapes, by assuming that Eq.~(\ref{SVAAcondition}) gives the sufficient conditions under which the SVAA gives accurate results for all smooth pulse shapes $A(t)$, where $\tau_{p}$ denotes the typical duration of the pulse. In Sec.~\ref{Sec:RWA} above Eq.~(\ref{doubleexRWA}) it was argued qualitatively that the RWA would be valid if the pulses are slow enough. For the validity of the SVAA we now find the more quantitative conditions~(\ref{SVAAcondition}). It will be checked in Sec.~\ref{Sec:numcomp} below that this condition also makes the RWA~(\ref{doubleexRWA}) valid for pulsed qubits, at least in combination with the standard RWA  condition $\Delta/\omega \ll 1$. After making both approximations, and by following the same reasoning as in Sec.~\ref{Sec:RWA}, the equations of motion assume the strikingly simple form
\begin{equation}\label{EqmotRWAandSVAA}
i \hbar \left( \begin{array}{c} \dot c_{1} \\ \dot c_{0} \end{array}\right) \simeq
\Delta  J_{n_{\rm res}}\left[ A(t)/\omega\right]\sigma_{x} \left( \begin{array}{c}  c_{1} \\  c_{0} \end{array}\right),
\end{equation}
valid for $\omega \gg \Delta,\tau_{p}^{-1}, \sqrt{A/\tau_{p}}$.
If the driving field would have been taken as proportional to $\cos(\omega t +\phi)$ instead of $\cos(\omega t)$, then this  nonzero driving phase $\phi$ at time $t=0$ would lead to phase factors $\exp(\pm i n \phi)$ that can be absorbed in a redefinition of $c_{0,1}$ in the same way as we dealt with the phase factors $\exp(\pm i n \chi)$ in Sec.~\ref{Sec:RWA}. Therefore, the equations of motion~(\ref{EqmotRWAandSVAA}) describe the approximated dynamics irrespective of the value of the static phase $\phi$. Our simple description would have lost some of its appeal if this had not been the case, since static phases are often hard to control experimentally.

The main results of this subsection are the approximate equations of motion~(\ref{EqmotRWAandSVAA}), in which the scaled amplitude $A/\omega$ should satisfy the weak condition~(\ref{SVAAcondition}), but otherwise is a free parameter. Thus the regime of strongly pulsed resonant driving is ready to be explored.

\subsection{Analytical solution of the approximated dynamics}
The coupled equations of motion~(\ref{EqmotRWAandSVAA}) obtained after RWA and SVAA allow a solution in closed form:
\begin{eqnarray}
c_{1}(t) & = &  -k_{1} \exp[i\, \Phi(t)] + k_{2}\exp[-i\, \Phi(t)],  \label{c1closed} \\
c_{0}(t) & = & k_{1} \exp[i\, \Phi(t)] + k_{2}\exp[-i\,\Phi(t)], \label{c0closed}
\end{eqnarray}
where $k_{1,2}$ are constants that are fixed by the state of the qubit at the initial time $t_{0}$.  Furthermore, in the exponents appears the dynamical phase factor
\begin{equation}\label{Phidef}
\Phi(t) = \Delta \int_{t_0}^{\,t}\mbox{d}\tau\,J_{n_{\rm res}}[A(\tau)/\omega].
\end{equation}
Unless stated otherwise, we will assume that the qubit starts in its ground state $|{0}\rangle$, in which case $k_{1,2}=1/2$ and the excited-state population becomes
\begin{equation}\label{sinphi}
P_{1}(t) = \sin^{2}[\Phi(t)]\qquad\mbox{for}\quad |{\psi(t_{0})}\rangle = |{0}\rangle.
\end{equation}
As we will see, this solution in closed form can be of considerable use. For example, in order to create desired final qubit states with the help of strong short pulses, one can employ Eqs.~(\ref{Phidef}-\ref{sinphi}) to optimize the pulse shapes $A(t)$, as will be explored in more depth in Section~\ref{pulseshaping}.

\section{Instantaneous coherent destruction of tunneling of a strongly pulsed qubit}\label{Sec:ICDT}

\begin{figure}
\centerline{\includegraphics{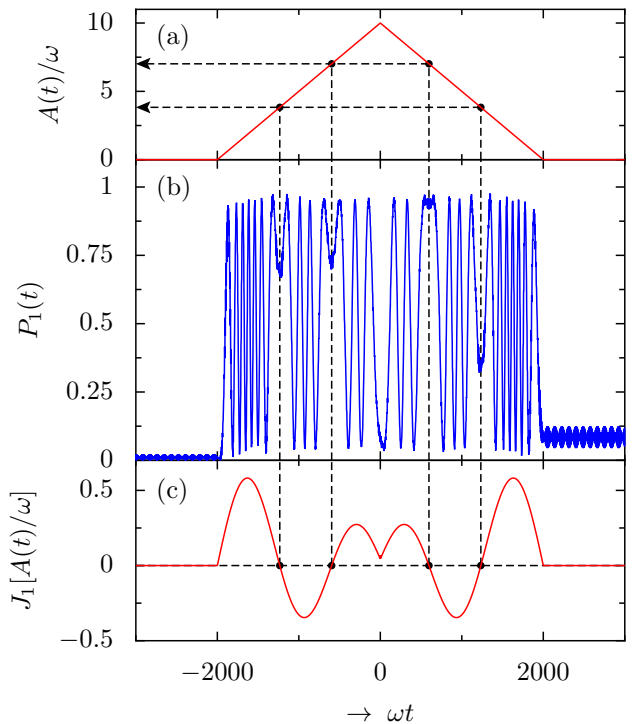}}\caption{Instantaneous coherent destruction of tunneling (ICDT) seen in a resonantly  ($\omega=\omega_{0}$) and strongly driven ($A/\omega=10$) qubit. The interaction is small, $\Delta/\omega = 0.075$, and the pulse width (FWHM) $\tau_{p} = 2000 \omega^{-1}$. Panel ({\bf a}) shows the  envelope function of the pulse with linear rise and fall, ({\bf b}) shows the numerically exact population dynamics of the initially (at $t_{0}=-3000\omega^{-1}$) unpopulated excited level $|1\rangle$, and ({\bf c}) depicts the Bessel function $J_{1}$ as a function of the scaled driving amplitude $A(t)/\omega$. A temporarily  frozen dynamics of the population around the dashed lines in (b) coincides with a zero of the Bessel function in (c), and the arrows in (a) point to $A(t)/\omega = 3.83$ and $7.02$, indeed the first two zeroes of $J_{- 1}$ (or of $J_{1}$). \label{figICDTfig}}
\end{figure}

In Sec.~\ref{Sec:RWA} it was anticipated that in strongly driven pulsed quantum bits a phenomenon called instantaneous coherent destruction of tunneling (ICDT) would occur. Here we will explore that further. In Figure~\ref{figICDTfig},
%%%%%%%%%%%%%%%%%%%%%%%%%%%%%%%%%%
%%%%%%%%%%%%%%%%%%%%%%%%%%%%%%%%%%
a pulse containing many resonant oscillations ($\omega = \omega_{0}$) rises linearly from zero to a maximum amplitude in the strong-coupling regime ($A/\omega \gg 1$) after which it falls back to zero with the same constant but now negative slope. In the weak-coupling regime, the oscillations of the qubit population would become faster as $A(t)$ grows, but the figure shows that this is certainly not the case during this stronger pulse. Rather, the rate of the oscillations is seen to correlate with the value of $J_{-1}[A(t)/\omega]$ (or, equivalently, with $-J_{1}[A(t)/\omega]$), which is in accordance with Eq.~(\ref{EqmotRWAandSVAA}).

In particular, at the times that this Bessel function vanishes,  the population dynamics is temporarily brought to a standstill. This lasts for only  short while, as the pulse amplitude keeps on changing, and the term instantaneous coherent destruction of tunneling  indeed seems appropriate for the phenomenon seen in Fig.~\ref{figICDTfig}. If one zooms in on the dynamics around an ICDT point, then one sees small fast oscillations remaining in the numerically exact dynamics. This suggests that the near-stationary term suppresses the effect of the fast oscillating terms, unless its amplitude nearly vanishes, in which case it becomes no more important than the neglected rotating terms.

For the linearly rising half of the pulse, the time integral over $h_{z}(t)$ is easy to do, and a term $A\cos(\omega t)/[\omega(\omega\tau_{p})] $ is neglected against $A(t)\sin(\omega t)/\omega$ when making the SVAA. From these formulae %and from Fig.~\ref{figICDT},
we again  see that this is all right to make the approximations in case $\omega\tau_{p}\gg 1$ and $(A/[\omega(\omega\tau_{p})]\ll 1$, which confirms the assumed generality of these conditions that were derived in Eq.~(\ref{SVAAcondition}) above for the two-sided exponential pulse.

It is to be expected and numerics confirms that as pulses become shorter while keeping the same maximum amplitude, instantaneous coherent destruction of tunneling may not be so clearly visible anymore as a temporarily frozen population. However, whether visible in the population or not, as long as the approximate description remains valid, the  effective interaction still vanishes at those instances  when the Bessel function $J_{n_{\rm res}}$ vanishes, and instantaneous coherent destruction of tunneling occurs.

\section{Numerical comparison of exact and approximated dynamics}\label{Sec:numcomp}

\subsection{Validity of RWA for approximation for two-sided exponential pulses}
\begin{figure}
\centerline{\includegraphics{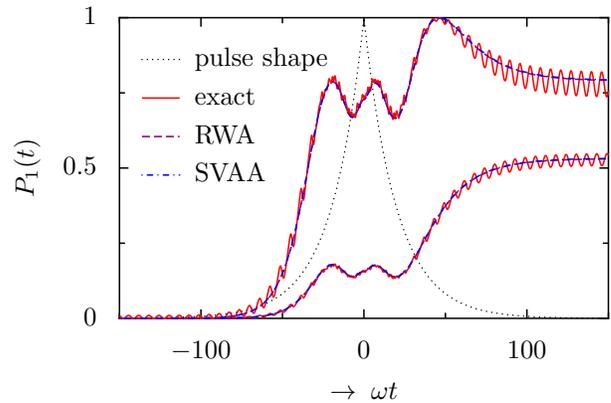}}\caption{Population dynamics $P_{1}(t)$ for a qubit resonantly ($\omega = \omega_{0}$) driven with a two-sided exponential pulse with maximum scaled amplitude $A/\omega = 10$ and FWHM width $\tau_{p} =40\omega^{-1}$, for $\Delta/\omega = 0.02$ (lower curves) and $\Delta/\omega = 0.05$ (upper curves). The red solid curve depicts the numerically exact dynamics of Eq.~(\ref{H2standardmatrix}), the purple dashed curve shows the dynamics after making the RWA of Eq.~(\ref{doubleexRWA}), and the blue dash-dotted curve represents the dynamics after also making the SVAA of Eq.~(\ref{sinphi}). The latter two curves  overlap. The dotted curve is the pulse form $p(t)$. \label{figcomparison}}
\end{figure}

We will now compare the exact dynamics of Eq.~(\ref{H2standardmatrix}) with the approximated dynamics. In Figure~\ref{figcomparison} a comparison is made for a qubit  driven by two-sided exponential pulses, for which we did a two-step approximation: first we made the RWA in Eq.~(\ref{doubleexRWA}), and then also the SVAA in Eq.~(\ref{EqmotRWAandSVAA}). The driving is quite strong but the approximated resonant dynamics closely follows the exact curves. The approximation is least accurate after the pulse, when the exact dynamics of the undriven qubit shows oscillations where the approximated population does not vary, and as expected these oscillations are larger for larger $\Delta$. As anticipated in Sec.~\ref{Sec:model}, making the SVAA hardly gives rise to extra inaccuracy once the RWA has been made, as exemplified by the overlap of the corresponding curves. Notice also in the figure that the discontinuity in the derivative of the pulse shape function at time $t=0$ does not lead to inaccuracies of the approximated dynamics after $t=0$.
%%%%%%%%%%%%%%%%%%%%%%%%%%%%%%%%%%
%%%%%%%%%%%%%%%%%%%%%%%%%%%%%%%%%%

\subsection{Validity of RWA and SVAA for gaussian pulses}
\begin{figure}
\centerline{\includegraphics{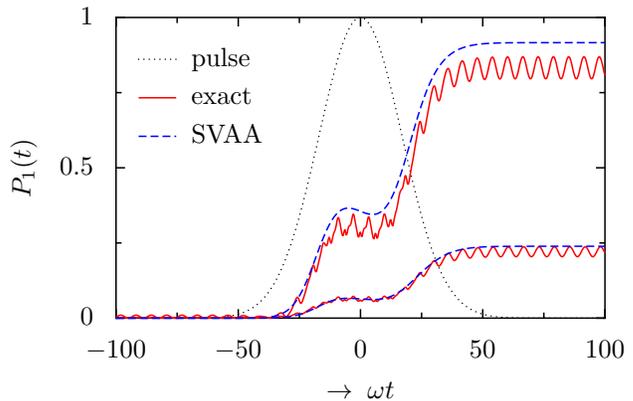}}\caption{Population dynamics $P_{1}(t)$ for a qubit resonantly ($\omega = \omega_{0}$) driven with a gaussian pulse with maximum scaled amplitude $A/\omega = 4$ and FWHM width $\tau_{p} =40\omega^{-1}$, for $\Delta/\omega = 0.02$ (lower curves) and $\Delta/\omega = 0.05$ (upper curves). The red solid curves depict the numerically exact dynamics of Eq.~(\ref{H2standardmatrix}),  while the blue dashed curve shows  the approximated dynamics after making both the RWA and SVAA as in Eq.~(\ref{EqmotRWAandSVAA}) and its solution~Eq.~(\ref{sinphi}). The dotted gaussian is the pulse form $p(t)$. \label{timegauss}}
\end{figure}

One can also compare the exact and the approximate dynamics for gaussian pulses  and this is exemplified in Figure~\ref{timegauss}. The exact and approximated curves for $\Delta/\omega = 0.05$ are not quite the same anymore, which shows that the condition $\Delta/\omega \ll 1$ for the RWA to hold must be observed rather stringently. For the smaller interaction $\Delta/\omega = 0.02$, the agreement between the two curves is good, and we stick to that small value for the interaction when discussing pulse shaping below.
%%%%%%%%%%%%%%%%%%%%%%%%%%%%%%%%%%
%%%%%%%%%%%%%%%%%%%%%%%%%%%%%%%%%%

\section{Pulse shaping for quantum state preparation with strong and short pulses}\label{pulseshaping}
\subsection{Weak-driving limit and area law}
The approximated final-state population depends on $\Phi(t)$, the integral given in Eq.~(\ref{Phidef}). This integral over a Bessel function with argument $A(t)/\omega$ can be simplified if the driving is weak, $A/\omega\ll 1$, because in general $J_{n}(x) \simeq x^n/(2^n n!)$ for $x\ll 1$, and also $J_{-n}(x) = (-1)^n J_{n}(x)$. For weak resonant driving with $\omega = \omega_{0}$, we have $n_{\rm res}=-1$, and the time integral then becomes an integral over the pulse area $\int_{t_{0}}^{t} \mbox{d}\tau A(\tau)$. In other words, for this specific resonance we find that the excitation probability at some time is given by the time integral over the pulse up to that time. Therefore, an intended final state can be engineered by the right choice of the total area under the driving pulse, where the precise form of the smooth pulse does not matter.   For other resonances $|n_{\rm res}|\ne 1$, the corresponding Bessel function has a nonlinear small-argument behavior, and the corresponding `laws' for weak driving are not  pulse area laws.

\subsection{Final states obtained by strong driving with a gaussian pulse}
\begin{figure}
\centerline{\includegraphics{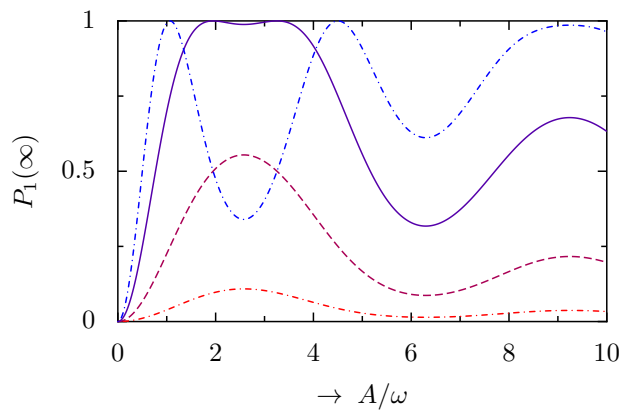}}\caption{Final excited-state probability $P_{1}(\infty)$ for a qubit resonantly driven ($\omega = \omega_{0})$ by a gaussian pulse, as a function of the scaled maximal pulse amplitude $A/\omega$, computed with the approximate solution Eq.~(\ref{sinphi}). Shown are curves corresponding to different pulse widths $\tau_{p}$ (FWHM): on the right, from bottom to top, the curves correspond to $\omega\tau_{p}=20$, $50$, $100$, and $150$. The interaction $\Delta$ equals $0.02\omega$ in all cases. \label{finalgauss} }
\end{figure}

Often the goal of driving a quantum system is to engineer an intended final state~\cite{Holthaus1992a,Wubs2007a,Fuchs2009a}, the pure excited state for example. If we consider a qubit driven by  gaussian pulses and take the interaction $\Delta$ small to keep the approximate treatment accurate -- and let us take $\Delta/\omega = 0.02$ -- then we have two parameters left to vary the pulse, namely the maximum amplitude $A$ and the pulse width $\tau_{p}$. Figure~\ref{finalgauss} depicts the probability to end up in the excited state as a function of the driving amplitude $A$, for several pulse widths.
%%%%%%%%%%%%%%%%%%%%%%%%%%%%%%%%%%
%%%%%%%%%%%%%%%%%%%%%%%%%%%%%%%%%%
From the figure it is clear that for the smallest pulse width $\omega \tau_{p}=20$ the probability $P_{1}(\infty)$ is small, whatever the driving amplitude. In order to reach complete inversion  $P_{1}(\infty)=1$, which according to Eq.~(\ref{sinphi}) corresponds to $\Phi(\infty) = \pi/2 + m\pi$ with integer $m$, one needs $\omega \tau_{p} =100$ or a bit less. Fig.~\ref{finalgauss} shows that small $\Delta$ and $\tau_{p}$ can not be compensated by huge driving amplitudes to obtain $P_{1}(\infty)=1$. For the longer pulses, the final probability becomes oscillatory as a function of the scaled driving amplitude $A/\omega$, and usually one would then choose the smallest value of $A$ to obtain the fully inverted final state. For shorter pulses, larger values of $A$ are needed. For large-amplitude driving the relation between $A$ and $\tau_{p}$ to obtain $P_{1}(\infty)=1$ is not precisely that of an area law. One can say that for strong driving the integral~(\ref{Phidef}) over the Bessel function replaces the area law~\cite{Holthaus1992a}.

In summary, a careful choice of the area under a pulse can be used to flip the state in the weak-driving regime for $n_{\rm res} =-1$. Choosing to satisfy the condition $\sin^{2}[\Phi(\infty)] = 1$ with $\Phi$ given in Eq.~(\ref{Phidef}) is a more general method that allows the  design of stronger pulses with arbitrary resonance order $n_{\rm res}$ that achieve a state flip of the qubit in a shorter time.

\section{Application: Spin qubit in diamond driven by a strong microwave pulse}
\label{Sec:diamond}

As an important application, here the Hamiltonian for a strongly driven NV center spin qubit in diamond will be briefly introduced, and two realizations of the standard CDT Hamiltonian with NV centers will be discussed, before discussing the theoretical results of previous sections in the light of the very recent experiments by Fuchs {\em et al.}~\cite{Fuchs2009a}.

The Hamiltonian $H_{0}$ of the undriven $S=1$ ground state of the NV center in a static magnetic field ${\bf B}$ is given by
\begin{equation}\label{H0}
H_{0} =  \Delta S_z^2 + g_{e}\, \mu_B {\bf B}\cdot {\bf S},
\end{equation}
which is the sum of a zero-field splitting and a Zeeman term.  The anisotropy term  $\Delta S_{z}^{2}$ describes a zero-field energy splitting of $\Delta = 2\pi \times 2.88\,\mbox{\rm GHz}$ between the $m=0$ and the $m=\pm 1$ states. The anisotropy implies a fixed quantization axis along the axis between the nitrogen and the vacancy, which will be called the $z$ axis. In the Zeeman term,  $g_{e}$ is the electron g-factor and $\mu_B$ the Bohr magneton. Magnitude and direction of the static magnetic field ${\bf B}=(B_{x}, B_{y}, B_{z})$ are not specified yet.
The NV center is driven by a strong microwave pulse
\begin{equation}\label{Vt}
V(t) = \sqrt{2} A(t) \cos(\omega t)\, {\bf n}_{\rm d}\cdot {\bf S}.
\end{equation}
The unit vector  ${\bf n}_{\rm d}$ describes which spin components of ${\bf S} = S_{x}\,\hat x + S_{y}\,\hat y + S_{z}\,\hat z$
the driving field couples to and how strongly. The factor $\sqrt{2}$ is taken out for later convenience. Without loss of generality, we can assume that ${\bf n}_{\rm d} = (\sin\theta, 0, \cos\theta)$, with $\theta$ the angle with the $z$-axis. The sum of Eqs.~(\ref{H0}) and~(\ref{Vt}) is the total Hamiltonian $H(t) = H_{0} + V(t)$ for the driven three-level atom.

Often the NV centers are operated such that one of the three levels can be neglected in the dynamics, and the remaining two levels constitute a long-lived qubit. For example, one can choose a static magnetic field ${\bf B}=(0, 0, B_{z})$ that lifts the $m=\pm 1$ degeneracy by an amount $\Delta E$ and a driving field with a frequency $\omega$ that is resonant with the $m=0 \leftrightarrow m=-1$ transition and an amplitude $A \ll \Delta E$. Then, if the system starts in the $m=0$ level, then typically the population of the $m=1$ level is negligible at all times. This is the situation in the recent experiments by Fuchs {\em et al.}~\cite{Fuchs2009a}. Notice that the energy $E_{m=0}=E_{0}=0$ for all field strengths $B_{z}$. After redefinition of the zero of energy, the effective two-level Hamiltonian is
\begin{equation}
H(t) = \varepsilon\, {\bm \sigma}_{z} + A(t) \cos(\omega t)[\sin\theta\, {\bm \sigma}_{x} + \cos\theta\, {\bm \sigma}_{z}],
\end{equation}
with $\varepsilon = |\Delta - g_{e} \mu_B B_{z}|/2$. In order to bring this Hamiltonian in the standard form of the CDT Hamiltonian, we rotate to a new basis where the Pauli matrices are $(\tau_{x}, \tau_{y}, \tau_{z})$ and where the driving pulse couples to $\tau_{z}$. In the new basis, the effective two-level Hamiltonian becomes (Case~1):
\begin{equation}\label{HdiamondCDT}
H(t) = \left( \begin{array}{cc} \varepsilon \cos \theta + A(t) \cos(\omega t) &  \varepsilon \sin \theta \\
 \varepsilon \sin \theta &  -\varepsilon \cos \theta - A(t) \cos(\omega t) \end{array} \right).
 %\qquad\mbox{(Case 1)}
\end{equation}
This Hamiltonian for a driven spin qubit in diamond is of the general form of Eq.~(\ref{CDTHamiltonianstandard}), with the transition frequency $\omega_{0}$ given by $|\varepsilon| \cos \theta =|\Delta - g_{e} \mu_B B_{z}|\cos (\theta)/2$ and the interaction $\Delta = \varepsilon \sin \theta$. The factor $\sqrt{2}$ in the driving amplitude canceled against the normalization factor of $S_{x}^{(1)}$. It is important to notice that the Hamiltonian~(\ref{HdiamondCDT}) can easily be changed upon variation of  external parameters. For example, the relative quantity $\Delta/\omega_{0} = \tan\theta$ is fully controlled by the angle of the driving interaction with the $z$-axis, and can assume all values. The  magnitudes of $\omega_{0}$ and $\Delta$ are both changed  upon variation of the external magnetic field $B_{z}$, while the strength $A$ of the driving field $V(t)$ is independently controlled by the input power.

Instead of driving the NV center off-axis, one could also drive it along the $z$-axis ($\theta = 0$), and apply a small static magnetic field $B_{x}$ in the $x$-direction. As before, a larger static $B_{z}$ field is to be applied to split the $m=\pm 1$ levels.  In this  case, the Hamiltonian after two-level approximation is (Case~2):
\begin{equation}
H(t) = \left( \begin{array}{cc} \varepsilon + A(t) \cos(\omega t) & g_{e} \mu_B B_{x} \\
g_{e} \mu_B B_{x} &  -\varepsilon - A(t) \cos(\omega t) \end{array} \right).
%\qquad \mbox{(Case 2)}
\end{equation}
Advantage of the second case may be that $\Delta = g_{e} \mu_B B_{x}$ and $\omega_{0} =\varepsilon = (\Delta - g_{e} \mu_B B_{z})/2$ can be varied independently, each by their own external parameter $B_{x,z}$. The Cases 1 and 2  are two ways to realize the standard CDT Hamiltonian for driven spin qubits in NV centers in diamond.

\subsection{Comparison with recent and proposals for future experiments}
Strongly pulsed driving of NV center spin qubits for Case 1 was reported very recently in Ref.~\cite{Fuchs2009a}, and for some strong and fast pulses the qubit changed state very rapidly, with possible ultrafast quantum state preparation applications. The experiments show that a great level of control can be achieved of the large-amplitude pulses that drive the long-lived NV center. The parameter regime was not the same as discussed here, mainly because the driving was chosen to make an angle $\cos\theta = \sqrt{1/3}$ with the $z$-axis, which according to Eq.~(\ref{HdiamondCDT}) means a strong static interaction in the CDT Hamiltonian with $\Delta/\omega_{0} = \sqrt{2}$. The approximate treatment adopted here is valid in the rather opposite $\Delta/\omega \ll 1$ limit. As a consequence, the short pulses that we found in this regime to invert the population are not quite as short as found in the experiments of Ref.~\cite{Fuchs2009a}. The advantage on the other hand of the $\Delta/\omega \ll 1$ regime is that it can be beneficial to design the right pulses with the help of analytical predictions for their effects on a qubit, such as presented here.

Another difference is that the current maximal experimental value of $A/\omega$ is $\simeq 1.1$. According to Fig.~\ref{finalgauss}, this enables one to invert qubit populations fast with pulses with $\omega \tau_{p} = 150$ and $\omega = 50\Delta$,  while some of the phenomena discussed above require even higher values of $A/\omega$. Thus the driving reported in Ref.~\cite{Fuchs2009a} is strong, but not quite as strong as would be needed to observe coherent destruction of tunneling.

It is therefore an interesting question whether (instantaneous) coherent destruction of tunneling could be observed in NV center spin qubits.  As shown in Fig.~\ref{figICDTfig}, the smallest value of $A/\omega$ for which CDT would occur for the $n_{\rm res} = -1$ resonance is $3.83$. Achieving this is challenging and would require  making $\omega_{0}$ smaller by increasing $B_{z}$ somewhat more, and by working with slightly higher pump powers at the lower resonant frequency. In Case 1 it would also require to drive at a small but nonzero angle with respect to the $z$ axis, so that $\tan\theta \ll 1$. In Case 2, again smaller resonance frequencies due to larger magnetic field $B_{z}$ and larger powers would be needed, in combination with a small static $B_{x}$-field. For fixed $A/\omega$, there is freedom to choose $A$ and $\omega$ such as to optimize the validity of the two-level approximation.
Similarly, in case of pulsed driving with $A/\omega$ somewhat increased and $\Delta/\omega$ reduced as compared to the recent experiments~\cite{Fuchs2009a}, the ICDT phenomena as discussed in this article are predicted to show up in strongly driven NV center qubits in diamond.

CDT by harmonic driving can serve as a useful method to lock a qubit with degenerate energy levels in a desired state, even in the presence of static couplings that otherwise would lead to tunneling.  For the NV center, the static magnetic field $B_z$ can be increased from the $850\,{\rm  G}$ of Ref.~\cite{Fuchs2009a} to $1023\,{\rm  G}$ (a slightly higher increase than what is needed above), so that $\varepsilon$ vanishes and the $m=0,-1$ levels become degenerate. In Case I, the tunnel coupling also vanishes with   $\varepsilon$. Unwanted static tunneling  terms $\Delta_{\rm noise}$ in the Hamiltonian could nevertheless lead to tunneling, but driving with parameters such that $\Delta_{\rm eff} = \Delta_{\rm noise} J_{0}(A/\omega)=0$ can prevent this. The smallest driving amplitude for which this occurs is $A = 2.40 \omega$. A first demonstration of quantum state preservation by CDT in NVs seems simpler in Case II, where the effective tunnel coupling $\Delta_{\rm eff}$ can be tuned both by the static magnetic field $B_x$ and by the driving.

Importantly from a quantum information perspective, the qubit state to be preserved  can be an arbitrary superposition, the coherence of which would not get lost while driving, at least on time scales that noise in the strong  driving is negligible.  This stabilization application of CDT is the opposite of fast manipulation. Of course it is equally important to be able to keep a qubit in a certain state as it is to be able to change its state, and both can be achieved by appropriately chosen strong high-frequency driving.

\section{Conclusions}
\label{Sec:conc}
In summary, the dynamics of qubits was studied that are resonantly driven by strong pulses. Starting with a simple pulse shape that allowed an analytical treatment, the parameter regime was explored in which the rotating-wave  and slowly-varying amplitude approximations are valid in case of strong resonant driving. In that regime, accurate solutions in closed form were presented for the strongly driven dynamics.  Coherent destruction of tunneling~\cite{Grossmann1991a,Grossmann1992a} was seen to be at work in the population dynamics of strongly pulsed qubits around specific instances in time, and this phenomenon could be called instantaneous coherent destruction of tunneling, or ICDT.

The closed solution of the approximate dynamics allows a simple method of pulse shaping in the large-amplitude driving regime, to prepare final quantum states using short resonant pulses. It was shown that the specific shape of the pulse is not important as long as one makes sure that a certain time integral~(\ref{Phidef}) has an intended value. The strong pulses that were were found to invert the qubit population fast can be seen as  generalizations of $\pi$-pulses for strong driving. In other words, this is another strong-driving method in the quantum state preparation toolbox, complementary to the novel method of Ref.~\cite{Fuchs2009a}. In specific applications, their merits  are to be compared with other methods to invert a spin, for example robust adiabatic passage schemes based on chirped pulses (used for calibration in~\cite{Fuchs2009a}).

Finally it was argued that the method of quantum state preservation by CDT and the pulse-shaping method discussed in this paper occur in a parameter regime that might soon be explored and used experimentally with  NV center spin qubits,  and there are good prospects of observing the phenomenon of instantaneous coherent destruction of tunneling.

\section*{Acknowledgments}
\noindent It is a pleasure to thank Peter H{\"a}nggi for being one of my Teachers in Science, and to wish him all the best for the future. The author acknowledges financial support by The Danish Research Council for Technology and Production Sciences  (FTP grant $\#274-07-0080$).

\appendix

\end{document}